\documentclass[12pt]{amsart}

\usepackage[margin=1in]{geometry}
\usepackage[lofdepth,lotdepth,caption=false]{subfig}
\usepackage{fancyhdr}
\usepackage{hyperref}
\usepackage{faktor}
\usepackage{amsmath, amssymb, graphicx}
\usepackage{mathtools}
\usepackage{xspace}
\usepackage{braket,placeins}
\usepackage{yfonts}
\usepackage{multicol}
\usepackage{multirow}
\usepackage{comment}
\usepackage{color}
\usepackage{setspace}
\usepackage{enumitem}
\usepackage{pst-node}
\usepackage{tikz-cd}
\usepackage{tikz}
\usepackage{tikz-network}
\usetikzlibrary{decorations.markings}
\usepackage{slashed}
\usetikzlibrary{calc}
\usepackage[mathscr]{euscript}
\usepackage[numbers]{natbib}
\usepackage[capitalize]{cleveref}

\newcommand{\cd}{\cdot}
\newcommand{\ra}{\rightarrow}
\newcommand{\pr}{\prime}

\newcommand{\C}{\mathbb{C}}

\newcommand{\R}{\mathbb{R}}

\newcommand{\bfJ}{\mathbf{J}}
\newcommand{\mfsu}{\mathfrak{su}}
\newcommand{\op}{\operatorname}

\newcommand{\id}{\mathrm{Id}}

\DeclareMathAlphabet{\mathpzc}{OT1}{pzc}{m}{it}

\newcommand{\mcL}{\mathcal{L}}

\newcommand{\mbbC}{\mathbb{C}}

\newcommand{\mbbR}{\mathbb{R}}

\theoremstyle{plain}

\theoremstyle{definition}
\newtheorem{dfn}{Definition}[section]
\theoremstyle{remark}
\newtheorem{rmk}{Remark}[section]

\newtheorem*{note}{Note}

\begin{document}

\title{The Smallest Interacting Universe}
\maketitle
\begin{center}
  \normalsize
  Modjtaba Shokrian Zini$^\dagger$, 
  Adam R. Brown$^*$, 
  Michael Freedman$^\bullet$
 \par   \bigskip

\let\thefootnote\relax\footnotetext{emails: \tt{v-modjtabas@microsoft.com, mr.adam.brown@gmail.com, michaelf@microsoft.com}}

\address{\textsuperscript{$\dagger$\label{3}}Research Consultant, Microsoft
}

 \address{\textsuperscript{*\label{1}}Google Research (Blueshift), Mountain View, CA 94043, USA, \&  \\
\indent \ \ Physics Department,
Stanford University, Stanford, CA 94305, USA
}

\address{\textsuperscript{$\bullet$\label{2}}
	Microsoft Research, Station Q, \& \\ Mathematics Department, UC Santa Barbara, CA 93106, USA 
}

\end{center}

\begin{abstract}

  The co-emergence of locality between the Hamiltonian and initial state of the universe is studied in a simple toy model. We hypothesize a fundamental loss functional for the combined Hamiltonian and quantum state, and then minimize this loss functional by gradient descent. We find that this minimization gives rise to a tensor product structure simultaneously respected by both the Hamiltonian and the state, suggesting that locality can emerge by a process analogous to spontaneous symmetry breaking.  We discuss the relevance of this program to the arrow of time problem.

In our toy model, we interpret the emergence of a tensor factorization as the appearance of individual degrees of freedom within a previously undifferentiated (raw) Hilbert space.   Earlier work \cite{freedman2021universe,freedman2021universeII}  looked  at the emergence of locality in Hamiltonians only, and in that context  found strong numerical confirmation of the hypothesis that raw Hilbert spaces of $\dim = n$ are unstable and prefer to settle  on tensor factorization when $n$ is not prime, expressing, for example, $n = pq$, and in \cite{freedman2021universeII} even primes were seen to ``factor'' after first shedding a small summand, e.g.\ $7 = 1 + 2 \cd 3$. This was found in the context of a rather general potential functional $F$ on the space of metrics $\{g_{ij}\}$ on $\mfsu(n)$, the Lie algebra of symmetries. This emergence of qunits through operator-level spontaneous symmetry breaking (SSB) may help us understand why the world seems to consist of myriad interacting degrees of freedom. But understanding why the universe has an initial Hamiltonian $H_0$ with a many-body structure is of limited conceptual value unless the initial state, $\ket{\psi_0}$, is also structured by this tensor decomposition. Here we adapt $F$ to become a functional on $\{g,\ket{\psi_0}\} = (\text{metrics}) \times (\text{initial states})$, and find SSB now produces a conspiracy between $g$ and $\ket{\psi_0}$, where they simultaneously attain low entropy by jointly settling on the same qubit decomposition. Extreme scaling of the computational problem has confined us to studying $\C^4$ breaking to $\C^2 \otimes \C^2$ and $\C^8$ breaking to $\C^2 \otimes \C^4$ or $\C^2 \otimes \C^2 \otimes \C^2$.\\

\end{abstract}
\thispagestyle{empty} 
\newpage 
\tableofcontents
\section{Introduction}
Mathematics has, from its beginning, flourished by asking the simplest possible questions. This is no less true today than in Euclid's time, although elaboration of language can, unfortunately, hide the simplicity. It is a measure of progress in physics (and indeed other sciences) that increasingly elementary questions can now be asked. Perhaps we are not yet ready to ask ``Why is there something?'', but this paper argues it is fruitful to ask ``Why is there more than one thing?''. And, in fact, our study suggests that this question is better phrased as ``Why does there \emph{appear} to be more than one thing?''. The \emph{inclination}, which is our subject, of $\C^4$ (or $\C^8$) to choose a specific \emph{breaking isomorphism} $\C^4 \cong \C^2 \otimes \C^2$ (or $\C^8 \cong \C^2 \otimes \C^4$ or $\C^2 \otimes \C^2 \otimes \C^2$) is a toy model for one thing appearing to have 2 (or 3) interacting degrees of freedom (dof), hence our title. We explore numerically how (operator level) spontaneous symmetry breaking (SSB) can produce a tiny interacting universe, say $\C^2 \otimes \C^2$, with a Hamiltonian $H_0$ and initial state $\ket{\psi_0}$---\emph{\textbf{s}imultaneously \textbf{k}nowing \textbf{a}bout \textbf{q}ubits} (\textbf{skaq}); In fact, knowing about the \emph{same} qubit decomposition.

In earlier work, some of us studied \textbf{kaq}, knowing about qubits, or related Majorana structures (\textbf{kam}), in the context of $H_0$ alone \cite{freedman2021universe,freedman2021universeII,freedman2021universeIII}. Here, we study the emergence of $(H_0,\ket{\psi_0})$, initial Hamiltonian and initial state, together. Hereafter we drop the subscript $0$ from $\psi_0$ to simplify expressions. It is very satisfying to see that they often conspire, jointly selecting a qubit structure w.r.t.\ which they both are nearly separable (disentangled). At the level of our toy model this matches the crudest feature of our universe, not only does it have things (plural) in it, but over 13.8 billion years it has gradually increased entropy at all scales. This patient metered increase in entropy requires the kind of conspiracy between $H_0$ and $\ket{\psi}$ seen in our numerics.

The `arrow of time' problem asks why we observe the universe to have such low entropy. In the account that has been standard since Boltzmann, the forwards-backwards time-asymmetry of our observed macroscopic world is to be explained not by a time-asymmetry in the underlying microscopic laws of physics, but instead by the time-asymmetry of the boundary conditions of the universe. In particular, the initial boundary condition, unlike the final boundary condition, is taken to have very low entropy. In this way the mystery of the time-asymmetry of the second law of thermodynamics is reformulated, and pushed backwards to be about understanding the special state in which our visible universe was born. However, understanding why the  universe started in a very special, low-entropy, state would give at best an incomplete understanding of the arrow of time. A low-entropy state evolving under the action of a generic Hamiltonian would not stay low-entropy for long: a single orthogonality time later it would have thermalized, the entropy would have reverted to a near-maximal value, and the  arrow of time would have ceased pointing. This is in stark contrast to the universe we see around us, which not only started in a very low entropy state, but is still in a (somewhat less) low entropy state despite in the intervening time having gone through perhaps $10^{120}$ orthogonal states. The reason our universe has taken so very long to thermalize is due to a special property not only of our state but also of our Hamiltonian. This special property includes, but is not limited to, spatial locality. We cannot have a full solution to the arrow of time problem until we have some principle that picks out not just the special initial state, but also the special Hamiltonian, and indeed that ensures the state and Hamiltonian are well-adapted to each other. The conspiracy we find between the Hamiltonian and the initial state is just the conspiracy we need to hope to fully address the arrow of time.

Our numerical work supports an understanding of only this most basic aspect of our reality: many interacting things with slowly growing sub-system entropies. The theory we are developing has not (yet) made contact with other properties of our universe, such as its spatial dimension (macroscopic or microscopic)---time is assumed as input as our starting point is the Schr\"{o}dinger equation. Nor with strings, foams, or fields. What we \emph{are} seeing is a robust mechanism (SSB) for a single, undifferentiated finite-dimensional Hilbert space to dress itself as an interacting world with an initial Hamiltonian and initial state both approximately respecting identical qubit decompositions, providing a mechanism for slowly increasing subsystem entanglement entropy $S(\op{tr}_A \psi_t)$ over time. It is pleasant to imagine redoing our calculation on a $10^{11}$ logical-qubit quantum computer and being able to empirically learn how the qunits we might find, if in the billions, would spatially organize. Might we see bits of Leech lattice emerging (a harbinger of bosonic string theory) with its bonds representing the strongest interactions?\footnote{\label{ftnote:gregmoore}We thank Greg Moore for the observation that the Leech lattice is a very natural bridge between local patterns of pairwise interaction and the 26 dimensions required in bosonic string theory for anomaly cancellation.} Unfortunately, this kind of large-scale brute force simulation is unlikely to be possible this century. We present the limited numerics that are currently feasible in the hope that it will inspire related models amenable to analytic methods. 

Let us take a step back and explain what we have studied. Let $H$ be an $n$-dimensional Hilbert space. In this paper $n=4$ or 8, but conceptually $n$ could be $n = 2^N$, and $N = 10^{100}$ in a black hole or early universe application. The restriction to powers of 2 is not essential \cite{freedman2021universeII} but may simplify the discussion. Here are the cast of actors: the Hilbert space $\C^n$ with $S^{2n-1}$ representing its unit sphere, the Hilbert space symmetries $\op{SU}(n)$, the infinitesimal symmetries, i.e.\ the Lie algebra $\mfsu(n)$, the moduli space $\mathcal{M}$ of unit norm $(\det = 1)$ metrics $g$ on $\mfsu(n)$ (this, of course, is also the space of left invariant Riemannian metrics on $\op{SU}(n)$ of fixed volume and also, after multiplication with $i$, the space of traceless Hermitian operators acting on $\C^n$), a functional (actually a family of functionals) $F_{24}: \mathcal{M} \times S^{2n-1} \ra \R$ described below, numerically detected local minima $(g,\ket{\psi})$ for $F_{24}$ ($g$ denotes a possible initial metric and $\ket{\psi}$ a possible initial state; $\ket{\psi}$ will shortly be used to build a source term for the integral $F_{24}$), a Gaussian probability distribution $G(g,\beta)$ on the unit volume metrics on $\op{Herm}_0(n)$, the space of traceless Hermitian operators on $\C^n$ (determined by the metric $g$ and an inverse temperature $\beta$), and finally a random Hamiltonian $H_0$ drawn from $\op{Herm}_0(n)$ with law $G(g,\beta)$. Our prologue on initial Hamiltonian and initial state concerns this pair $(H_0, \ket{\psi})$. In summary, the cast is:
$$
	(\C^n, S^{2n-1}),\ \op{SU}(n),\ \mfsu(n),\ \mathcal{M},\ F_{24}: \mathcal{M} \times S^{2n-1} \ra \R,\ (g,\ket{\psi}),\ G(g,\beta), \text{ and } (H_0,\ket{\psi})
$$

Papers \cite{freedman2021universe,freedman2021universeII,freedman2021universeIII} followed a similar plot but with $\ket{\psi}$ omitted. There, \textbf{kaq} metrics $g$ were found while in this paper \textbf{skaq} pairs $(g,\ket{\psi})$ are found. Let us run through the connective steps in the above chain. First, putting a metric on $\mfsu(n)$ amounts to distinguishing between easy vs.\ hard (sometimes expressed as cheap vs.\ expensive) variations in the (pure) state of a quantum system. This has been extensively studied in two, now highly overlapped, communities, quantum computing (QC) and quantum gravity (QG). Briefly, easy in QC means a reproducible controllable interaction, and in QG is a strong interaction between event horizon dof \cite{nielsen2005geometric,brown2017quantum}. The concept of many-body physics can be well captured by the choice of a metric $g$ and its induced probability distribution $G(g,\beta)$ which tells you what kind of interactions are likely to appear in a Hamiltonian $H$ and which are far-fetched.

As we noted, $\sqrt{-1} \times \mfsu(n) = \op{Herm}_0(n)$. That is, up to a factor of $i$, $\mfsu(n) =\{$traceless skew-Hermitians$\}$ and $\op{Herm}_0(n)$ agree so $g$ can at any moment be regarded as a metric on the Lie algebra, and therefore a left invariant metric on $\op{SU}(n)$, or as a metric on the space of traceless Hamiltonians. We go back and forth between these pictures. 

Thinking of $g$ as a metric on traceless Hermitians; for $\beta > 0$ define the probability distribution
\[
	G(g,\beta) \coloneqq \frac{1}{Z(\beta)} e^{-\beta\langle H, H\rangle_g} = \frac{1}{Z(\beta)} e^{-\beta g_{ij}h^i h^j}, \text{ where } H = h^i H_i.
\]
where $h^i \in \mbbR$ and $\{H_i\}_{i=1}^{4^n-1}$ are the basis elements of $\op{Herm}_0(n)$. Then the $H_0$ above (in the pair $(H_0,\ket{\psi})$) is to be drawn randomly from such a distribution, determined by the $g$ component of the local minimum $(g,\ket{\psi})$. Note that in the low temperature limit, $\beta \ra \infty$, $H_0$ concentrates on the cheapest ``eigen-direction'' or principal axis of $g$. The conspiracy referenced above that $(g,\ket{\psi})$ are \textbf{s}imultaneously \textbf{kaq} is essentially the statement that all the principal axes of $g$, \emph{and} $\ket{\psi}$ have extremely low entanglement entropy within their respective representation w.r.t\ a single qubit decomposition of $\C^n$. $H_0$ transforms as an operator on $\C^n$ and naively $\ket{\psi}$ does so as a vector, however it is convenient to promote $\ket{\psi}$ to also be an operator in the usual way: $\ket{\psi} \mapsto J:= i|\psi\rangle \langle \psi|$. With $H_0$ and $J$ now on the same footing, we define (\cref{ssec:defnF24}) functionals $F_{24}$ using $J$ as a source term. The striking numerical finding is that local minima $(g,\ket{\psi})$ for $F_{24}$ quite often exhibit \textbf{skaq} structures. The extent to which many local minima $(g, \ket{\psi})$ ``conspire'' to select a qubit structure of low entropy for both $g$ and $\ket{\psi}$ cannot possibly be due to chance, see \cref{sec:summary}.

There are locals in mathematics and physics where rather general families of functionals (see \cite{cohn2019universal} in the context of sphere packing) lead to minima with rigid geometrical properties. We believe we have found another such local and call the present instance \textit{metric-state-crystallization}. Given $F_{24}$, $(g,\ket{\psi})$ is the crystal in this analogy. It arises via spontaneous symmetry breaking (SSB), but now acting on the level of operators, not states.

The final piece of this sketch is to describe the functional $F_{24}$. Without a source term functionals on the metric alone can sometimes have a direct geometric interpretation, e.g.\ Ricci scalar curvature (investigated in \cite{freedman2021universe,freedman2021universeII}). With the source term, we know only an algebraic/analytic interpretation via path integrals.

The Lie algebra $\mfsu(n)$ is entirely captured by a 3-tensor $c_{ij}^k$, the structure constants, $[H_i,H_j] = c_{ij}^k H_k$. A metric $g_{ij}$ is a 2-tensor on $\mfsu(n)$, a source $J=i\ket{\psi}\bra{\psi}$ acts as a covector (functional) on $\mfsu(n)$ via the $L^2$ (Killing form) inner product. So, the obvious functionals would be closed Penrose diagrams made from these three tensors: $c,g,J$. We know, both analytically and from numerical experiments, that individual diagrams are convex functions on $\mfsu(n)$ \cite[App. C]{freedman2021universe}, so not interesting for SSB. But both finite and infinite combinations of closed diagrams can be extremely interesting and we study these. But which to choose? We settled here on the euclidean version of the simplest perturbed Gaussian integral that can be invariantly constructed from $(g,c,J) = (\text{quadratic term, cubic interaction, source})$. As expected with Gaussian integrals, we will have a couple of adjustable expansion parameters, but our functional $F_{24}$ belongs to the simplest possible family of perturbed Gaussians constructed from the tensors $c$, $g$, and $J$ (where perturbation is not zeroed out by the symmetry $c_{ij}^k = -c_{ji}^k$). In practice we only use the first 9 diagrams with either 2 or 4 copies of $c$---hence the notation $F_{24}$. We follow the standard physical practice of truncating a perturbative expansion to its early terms. We have no rigorous results on what such a procedure actually means, but empirically we find functions whose minima $(g,\ket{\psi})$ have this extraordinary \textbf{skaq} property. We believe that similarly structured minima will generically be found for a wide range of even less well-motivated combinations of Penrose diagrams. This intuition comes from examining local minima found while our program still suffered from bugs and evaluated wrongly included diagrams.

To summarize, we generate some interesting looking functions $F_{24}$ on metrics $\times$ pure states ($\mathcal{M} \times S^{2n-1}$), and then do gradient descent using algorithms in the machine learning literature to find local minima. When we dive in and examine these minima we often see an extraordinary (and impossible to account for by chance) conspiracy to lower the entropy of both $g$ and $\ket{\psi}$ simultaneously in some autonomously selected preferred tensor structure $\C^4 \cong \C^2 \otimes \C^2$ or $\C^8 \cong \C^2 \otimes \C^4$ or $\C^8 \cong \C^2 \otimes \C^2 \otimes \C^2$. 

The initial state and Hamiltonian of the observed universe both have a very special structure. In this paper we try to understand general generative principles that could naturally give rise to this structure. One could imagine trying to break this problem into two steps, first producing a Hamiltonian, and then from that Hamiltonian extracting a state, for example the ground state, or in a gravitational context the Hartle-Hawking state \cite{Hartle:1983ai}. But the generative principle explored in this paper instead generates them together, in a single step. The two components---state and Hamiltonian---are treated on an equal footing, and the minimization is  performed over both simultaneously. 

Let's now turn to giving some precise definitions, then we will give a high level summary of findings, and then discuss our  data and techniques.

\subsection{Definition of the functional family $F_{24}$}\label{ssec:defnF24}~\\
\indent We would like to build a family of perturbed Gaussian integrals over $\mfsu(n)$ or equivalently the traceless Hermitian matrices $\op{Herm}_0(n)$. Symmetry consideration (see \cite{freedman2021universe}) kills the cubic term (or the quadratic) in the most obvious implementation using Bosonic (or Fermionic) variables, so as in \cite{freedman2021universe,freedman2021universeII,freedman2021universeIII} we adopt the expedient of integrating over three copies of $\op{Herm}_0$. This is our integral with parameters $k$ and $k_J$:
\[
	F_{24}(g,J,k,k_J) :\approx \int_{\vec{x} \in \R^{3(4^n-1)}} d\vec{x}\ \exp\left[-kG_{IJ}x^Ix^J + c_{ijk} y_1^iy_2^jy_3^k + k_J \mathbf{J} \cd x\right]
\]
where the subscript for $F$ reminds us that we keep only the terms of order 2 and 4 in $c$ in the perturbative expansion. In detail, the notation borrowed from \cite{freedman2021universe} is as follows: $x = (y_1,y_2,y_3)$ with $y_o \in \R^{4^n-1}, o \in \{1,2,3\}$, $I = (i,o)$ and $x^I = y_{o}^i \in \mathbb{R}$, and finally $G_{IJ}x^I x^J = g_{ij}y_1^i y_1^j + g_{ij} y_2^i y_2^j + g_{ij} y_3^i y_3^j$, i.e. $G = \begin{pmatrix}
    g & 0 & 0 \\
    0 & g & 0 \\
    0 & 0 & g
\end{pmatrix}$. The structure constants $c_{ij}^k$ of the Lie algebra are
\begin{equation}\label{c_ijkdfn}
    [y_i, y_j] = c_{ij}^k y_k \text{ and } c_{ijk} = c_{ij}^{k^\pr} g_{k^\pr k}.
\end{equation}

Regarding the new notations, in addition to the new parameter $k_J$ which regulates the impact of the source, we have $\mathbf{J}\cdot x = \mathbf{J}_{i,1}\cdot x^{i,1}+\mathbf{J}_{i,2}\cdot x^{i,2}+\mathbf{J}_{i,3}\cdot x^{i,3}$ with $\mathbf{J}_{i,o}$ being copies of $J=i\ket{\psi}\bra{\psi}$ projected onto $\mfsu(n)$, in other words $\mathbf{J}_{i,o} = J - \frac{i\op{Id}_{\op{Herm}(n)}}{4^n}$.

To derive the relevant diagrams, we follow the same procedure outlined in \cite[Eqs. (10-15)]{freedman2021universe}, with help from \cite[Appendix]{bar1997aarhus} due to the inclusion of the source term. The perturbative series expansion leads to 
\begin{align}
    \Big[ (c_{ijk} \frac{\partial}{\partial V_1^i}\frac{\partial}{\partial V_2^j}\frac{\partial}{\partial V_3^k} + k_J(\mathbf{J}_{i,1}\frac{\partial}{\partial V_1^i}+ \mathbf{J}_{j,2}\frac{\partial}{\partial V_2^j}+ \mathbf{J}_{k,3}\frac{\partial}{\partial V_3^k}))^m (-kG^{IJ}V_IV_J)^{l}\Big] \Big|_{\vec{V}=0}.
\end{align}
The variable $V$ is a dummy variable used to compute the Gaussian integral. It should not be confused with $x$ or $y$ even though it has a similar indexing notation. To understand which tensor diagrams emerge from this expansion, notice that the pairing between $i,j,k$ with $1,2,3$ and the block-diagonal structure of $G$ plays a significant role in the simplification of this expression \cite{freedman2021universe}. 

Recall that we are expanding the expression up to the fourth power of $m$. To get a closed diagram, one needs to exactly pair the number of differentials, such as $\frac{\partial}{\partial V_{1}^i}$, picked from the first expression which has power $m$, with the variables, such as $V_1^{i}$, picked from the second expression which has power $l$. Call $v_c,v_\mathbf{J}$ the nonnegative number of $c, \mathbf{J}$ vertices in any diagram. One can easily observe the following two equations: $v_c + v_\bfJ = m$ and $3v_c + v_\bfJ = 2l$. The second equation implies that $v_c \equiv v_\bfJ \pmod 2$, and thus $m \equiv 0 \pmod 2$ from the first equation. Solving the equations yields $v_c = l - m/2$ and $v_\bfJ = 3m/2 - l$, and since both are nonnegative numbers, if $m=2$, we have $1 \le l \le 3$, and $1 \le l \le 6$ when $m=4$. The rest is case-checking and we find that only the following pairs give diagrams with a nonzero coefficient $(m,l) = (2,3),(2,1),(4,6),(4,4),(4,3),(4,2)$.

Next, we describe each diagram, some of which are from \cite{freedman2021universe}. Note that we will use the vertex $c$ to refer to structure constants $c_{ij}^k$, and not $c_{ijk}$. This distinction implies the presence or absence of certain factors of $g^{\pm 1}$ as explained later for each diagram. In all diagrams, edges of color black, red, and green imply pairing through $g^{-1},g$, and $\op{Id}$ respectively.

For $m=2,l=3$, we obtain the theta diagram in \cref{Thetadiagram} which we borrow from \cite[Fig. 4]{freedman2021universe}. For $m=2,l=1$, we essentially get $\bfJ\cdot G \bfJ$ and call it the bar diagram in \cref{Bardiagram}. $m=4,l=6$ gives the tetrahedron, tincan and the double theta diagrams as shown in \cite[Fig. 5]{freedman2021universe}. For $m=4,l=4$ we have the theta diagram joined by the bar diagram, and a new diagram shown in \cref{thetaJsplit}. For $m=4,l=3$, we get the new diagram in \cref{forkdiagram}. Finally, $m=4,l=2$ gives two copies of the bar diagram.

\begin{figure}[h]
    \centering
\begin{tikzpicture}
\Vertex[x=1,label=$c$]{A}
\Vertex[x=1,y=-2,label=$c$]{B}
\Edge[label=$k$, color = red](A)(B)
\Edge[bend=65,label=$i$](A)(B)
\Edge[bend=-65,label =$j$](A)(B)
\end{tikzpicture}
    \caption{\small{Theta diagram. All diagrams are trivalent networks without any loop, and vertices are the structure constants $c_{ij}^k$. Each vertex has indices $i,j,k$ which are paired with their counterpart in another vertex. This pairing is done using $g$ along edge of type $k$ (colored red) and $g^{-1}$ for type $i$ and $j$. Notice the  $g$ factor comes from the lowering of the $k$ index in \cref{c_ijkdfn} (two such factors from the two $c_{ijk}$, and one $g^{-1}$ factor from $G^{IJ}$).}}
    \label{Thetadiagram}
\end{figure}
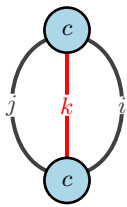

\begin{figure}[h]
    \centering
    \begin{tikzpicture}
    \Vertex[x=1,label=$\bfJ$]{A}
    \Vertex[x=1,y=-2,label=$\bfJ$]{B}
    \Edge[label=$k$](A)(B)
    \Vertex[x=1,label=$\bfJ$]{A}
    \Vertex[x=1,y=-2,label=$\bfJ$]{B}
    \Edge[label=$k$](A)(B)
    \end{tikzpicture}
    \caption{\small{Bar diagram. Notice the pairing is done via $g^{-1}$ and not $g$, as unlike the theta diagram where there is a factor of $g$ from the $k$ index of $c_{ijk}$, there is no such factor from $\bf{J}_{k,3}$. We have the same pairings for indices $i$ and $j$, and note that their values are the same as the one above.}}
    \label{Bardiagram}
\end{figure}
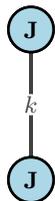

\begin{figure}[h]
    \centering
    \begin{tikzpicture}
    \begin{scope}[shift={(-4,0.0)}]
    \Vertex[x=1,label=$c$]{A}
    \Vertex[x=1,y=-2,label=$c$]{B}
    \Vertex[x=-1, y=-1.5,label=$\bfJ$]{C}
    \Vertex[x=-1, y=-0.5,label=$\bfJ$]{D}
    \Edge[label=$k$, color = red](A)(B)
    \Edge[bend=65,label=$j$](A)(B)
    \Edge[label =$i$](C)(B)
    \Edge[label =$i$](D)(A)
    \end{scope}
    \Vertex[x=1,label=$c$]{A}
    \Vertex[x=1,y=-2,label=$c$]{B}
    \Vertex[x=-1, y=-1.5,label=$\bfJ$]{C}
    \Vertex[x=-1, y=-0.5,label=$\bfJ$]{D}
    \Edge[label=$k$, color = red](A)(B)
    \Edge[bend=65,label=$i$](A)(B)
    \Edge[label =$j$](C)(B)
    \Edge[label =$j$](D)(A)
    \begin{scope}[shift={(+4,0.0)}]
    \Vertex[x=1,label=$c$]{A}
    \Vertex[x=1,y=-2,label=$c$]{B}
    \Vertex[x=-1, y=-1.5,label=$\bfJ$]{C}
    \Vertex[x=-1, y=-0.5,label=$\bfJ$]{D}
    \Edge[label=$j$](A)(B)
    \Edge[bend=65,label=$i$](A)(B)
    \Edge[label =$k$, color=green](C)(B)
    \Edge[label =$k$, color=green](D)(A)
    \end{scope}
    \end{tikzpicture}
    \caption{\small{Theta-J diagram. The first two on the left give same value due to the symmetry of $i,j$ pairings. In the instance on the right, since we use the $k$ index in pairing the two copies of $\bfJ$ to $c$, then one factor of $g$ from $c^k_{ij} \to c_{ijk}$ is cancelled by the $g^{-1}$ factor from $G^{IJ}$, thus giving an identity pairing between $\bf{J}$ and $c$, and hence the color green.}}
    \label{thetaJsplit}
\end{figure}
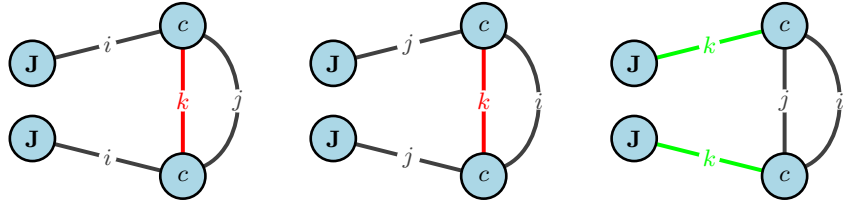

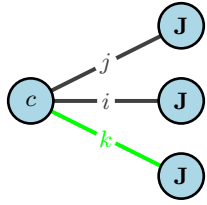
\begin{figure}[h]
    \centering
    \begin{tikzpicture}
    \Vertex[x=1,label=$c$]{A}
    \Vertex[x=3,y=-1,label=$\bfJ$]{B}
    \Vertex[x=3,label=$\bfJ$]{C}
    \Vertex[x=3, y=1,label=$\bfJ$]{D}
    \Edge[label=$k$, color = green](A)(B)
    \Edge[label=$i$](A)(C)
    \Edge[label =$j$](A)(D)
    \end{tikzpicture}
    \caption{\small{Fork diagram. $\bfJ$ is paired through $G$ with $c$, and the $k$ index becomes an identity pairing similar to \cref{thetaJsplit}.}}
    \label{forkdiagram}
\end{figure}

\subsection{Definition of \textbf{skaq}}~\\
\indent In earlier work \cite{freedman2021universe,freedman2021universeII} knows about qubits (\textbf{kaq}), or in \cite{freedman2021universeIII}, knows about Majoranas (\textbf{kam}), was the key concept. Here we study \textbf{s}imultaneously \textbf{kaq} (\textbf{skaq}), meaning a simultaneous factorization of the principal axes (compared to the Killing form) of the metric $g$ \emph{and} the initial state $J$.

It is easily proven by induction that if $n = p_1 \cdots p_l$ is a prime factorization then $\op{Herm}(n) = \op{Herm}(p_1) \otimes \op{Herm}(p_2) \otimes \cdots \otimes \op{Herm}(p_l)$. Note we have temporarily dropped the traceless condition, and use the natural inclusion $\op{Herm}_0(n) \subseteq \op{Herm}(n)$ below to rectify this.

\begin{dfn}[$\text{\cite[Def. 2.1]{freedman2021universeII}}$]\label{dfn:qunitstructure}
	A qunit structure on $\C^n$ is an equivalence class of $\ast$-isomorphisms\footnote{We note that we have changed the notation in the reference \cite[Def. 2.1]{freedman2021universeII} to avoid overburdening the notation $J,j$ in our paper.} $\mathbf{Q}: \C^{p_1} \otimes \cdots \otimes \C^{p_l} \xrightarrow{\cong} \C^n$ where  two are equivalent if related by the left action on the factors by $\mathrm{U}(p_1) \times \cdots \times \mathrm{U}(p_l)$. Thus qunit structures are parameterized by $\mathrm{U}(p_1) \times \cdots \times \mathrm{U}(p_l) \setminus \mathrm{U}(n)$. Note that $\mathbf{Q}$ induces an isomorphism $\mathbf{q}: \op{Herm}(p_1) \otimes \cdots \otimes \op{Herm}(p_l) \xrightarrow{\cong} \op{Herm}(n)$.
\end{dfn}
Below is a modification of \cite[Def. 2.2]{freedman2021universeII} for our purpose:
\begin{dfn}\label{dfn:skaq}
	A $(\text{metric } g_{ij}, \text{state } J)$ on $\op{Herm}(n) \times S^{2n-1}$ is \textbf{skaq} iff $g$ is \emph{not} ad-invariant, yet there is an isomorphism $\mathbf{q}$ (induced from $\mathbf{Q}$ above) so that $g_{ij}$ possesses a complete set of $n^2 - 1$ principal axes $\{H_k\}_{1 \leq k \leq n^2-1}$ with
	\[
		H_k = \mathbf{q}(H_{1,k} \otimes \cdots \otimes H_{l,k}) \text{ where } H_{s,k} \in \op{Herm}(p_s),\ 1 \leq s \leq l
	\]
	and
	\[
		J = \mathbf{q}(J_1 \otimes \cdots \otimes J_l), \text{ where } J_s,\ 1 \leq s \leq l, \text{ is a rank one Hermitian projector in } \C^{p_s}.
	\]

	\noindent In other words, $\C^n$ admits a tensor structure so that the principal axes of $g_{ij}$ ($=$ eigenvectors of $g_i^j$ where the Killing form is used to raise the index) \textit{and} $J$ all have compatible tensor structures. Note that $H_k \in \op{Herm}_0(n)$, but $H_{s,k} \in \op{Herm}(p_s)$.
\end{dfn}

\begin{note}
	In earlier papers \emph{partially \textbf{kaq}} was used when $p_1, \dots, p_l$ was a nontrivial factorization but not necessarily into primes. We encounter this circumstance again in our numerics, corresponding to $8 = 2 \times 4$ and use the corresponding term \emph{partially \textbf{skaq}}.
\end{note}

Before proceeding to the executive summary of our findings, we note that the question of why numerics have been restricted here to $n\le 8$ has been answered in \cite[Sec. 3.1.3]{freedman2021universeII}, where the issue of diagram contraction, their storage requirement, and the unparallelizable nature of the procedure are the main bottlenecks.

\section{Summary of results}\label{sec:summary}

The definitions of (partial)-\textbf{kaq} and (partial)-\textbf{skaq} are mathematical idealizations, which correspond to data landing on a real analytic variety of high codimension, and as such it can not be verified to infinite precision in numerical experiments. Moreover, the most basic ``facts'' of slow and steady entropy growth in our universe, reproduced by our tiny toy models, is not compatible with exact \textbf{skaq} (or partially \textbf{skaq}) minima. Indeed, if these conditions are mathematically exact, then the Schr\"{o}dinger evolution will maintain the initial disentangled condition of the initial state. In assessing how close to these conditions we are, there are two very natural measures, which are fairly close in the case of $\mfsu(4)$ and very different in the case of $\mfsu(8)$. The most straightforward measure is just to use entropy scores directly; For example, applied to principal axes with the highest entropy $H_{i}$ of $g$, or a state vector $\ket{\psi}$, we observe that one state has only 10\% of the entropy of another, $S(\ket{\psi_1})=0.1S(\ket{\psi_2})$, and use this as part of a score to determine \textbf{skaq}ness. A more revealing approach is to compute, or numerically simulate (which is what we have done), the probability density function of $S$ on Gaussian random states (or appropriate classes of metrics) and score \textbf{skaq}ness based on the rareness of the values derived from local minima. Depending on the codimension of the corresponding varieties, the difference can be a factor of 3 or so for entropy of vectors in $\mbbC^2 \otimes \mbbC^2$ (\cref{fig:su4J}) or a factor of $10^3$ for $\mbbC^2 \otimes \mbbC^4$ as low entropy states in this decomposition are much rarer (\cref{fig:su8J}). This density approach, applied to our data sets (225 $\mfsu(4)$ local minima and 282 $\mfsu(8)$ local minima), show very pronounced global patterns- spikes of various entropy distributions near zero which could not possibly arise by chance by sampling the generic ensembles. To single out a typical example of one of our conclusions and its statistical basis, glance ahead to \cref{fig:su4J}. On the left, we show a histogram of entropies for a Gaussian random ensemble of 100,000 unit vectors in $\C^2 \otimes \C^2$, and on the right the histogram of the entropies of the 225 state vectors extracted from the 225 local minima we found studying $\mfsu(4)$.  It seems  superfluous to quantify our confidence that our data is not drawn randomly from the former distribution. If one inputs the mean and variance of the random distribution, it is an 11 sigma event for a sample of 225 to have mean as small or smaller than our data, 0.4024.  In fact the structure of the data, the huge spike near zero, make it even less likely to have arisen by chance from the baseline ensemble. 

Throughout the paper we have tried to find the most relevant and ``fairest'' comparison ensembles for our data sets. We will not give (additional) quantitative measures of confidence as they are always enormous if one accepts the comparison ensemble.  It is the latter point where some degree of judgement must enter. The comparison ensemble for state vectors (as above) does not allow much ambiguity, but it is a bit more subtle for metrics. For numerical stability during gradient descent of a metric, we start from rather small variance (0.1) distributions to find local minima. We used various heuristics, to select appropriate corresponding variances for the comparison ensembles (see \cref{sec:genrandombaselines}). Typical entropies  of the comparison ensembles do shift about with these choices, so there is not a unique baseline from which to generate a `null hypothesis'. But in all cases, the argument we make rests on the presence of a sharp low entropy spike, not present in any reasonable comparison ensemble. In fact, quite often the data from local minima yields a distribution quite similar to the comparison distribution except for a pronounced spike near zero in the data distribution (see \cref{fig:su8J} for a particularly clear example.)

Adding the additional requirement of a small $S(\ket{\psi})$, i.e.~going from \textbf{kaq} to \textbf{skaq}, does require us to somewhat loosen the criteria used in the earlier \cite{freedman2021universeII} in order to have a few exemplars to look at in detail. As we have said, this relaxation should not raise doubts about the statistical significance of our findings---that case is made from ensemble entropy distributions. Generally, we compute up to 8 floating point decimal places and trust our numerics at least to the first three places. In \cite{freedman2021universeII}, \textbf{kaq} (or partial-\textbf{kaq}) was declared if the maximum entropy of principal axes (out of the 15 for $\mfsu(4)$ and 63 for $\mfsu(8)$) was below $<0.001$. Now as we move, in this paper, to the study of \textbf{skaq}, our strictest criteria is that we should see, in the same tensor decomposition, entropy lower than the 5\% w.r.t. random ensembles for $J$. For the test on $g$, we consider those $g$'s for which the maximum entropy of principal axes is low enough that they are not even seen in the random baselines: 2e-3 for $\mfsu(4)$, and 1.1e-2 and 3e-2 for partial and full decomposition for $\mfsu(8)$, respectively. For $\mfsu(4)$, we find 3 out of the 225 local minima and for $\mfsu(8)$, there are 6 partial-\textbf{skaq} out of 282. But the most striking findings are the how different the overall entropy statistics are at local minima compared to a random baseline.

Very broadly there are three findings, each is witnessed by entropy distributions derived from the set of $F_{24}$-local minima $(g,\ket{\psi})$, strikingly different from the natural random ensembles. Over $\mfsu(4)$, \textbf{skaq} structure is evidenced, when all local minima are considered together, distributions of entropic measures exhibit large spikes near zero, whereas random ensembles have vanishing density near zero, that is near $S(g)=0$, and $S(\ket{\psi})=0$. This is our version of the first theorem of mathematics: $2\times 2=4$. Please glance forward to the two pairs of \cref{fig:su4g,fig:su4J} contrasting the entropy of the most entangled principal axes $H_i$ and the state $\ket{\psi}$ within our data set, with that of the random baselines. Over $\mfsu(8)$, where there are many more constraints, \textbf{skaq} metrics, w.r.t all their parameters, are now rare. However, considering the entire sample of local minima $(g,\ket{\psi})$ we see a very similar double spike near zero entropy singling out a choice of decomposition $\mbbC^8 = \mbbC^2 \otimes \mbbC^4$; our version of $2\times4=8$. Again, please look ahead to figures \cref{fig:su8g,fig:su8J} to see the comparison of entropy distributions.

Our third finding came as a surprise and we invite the reader to join us in trying to make sense of it. As the previous two findings, it is too significant to be a fluke. On $\mfsu(8)$, we find a decided \textit{resistance} to a full \textbf{skaq} decomposition of $\mbbC^8$ into $\mbbC^2 \otimes \mbbC^2 \otimes \mbbC^2$. Looking back to the data from \cite{freedman2021universeII}, we see that without the $J$ term ($k_J$ = 0), no full \textbf{kaq} decomposition were found for $\mfsu(8)$ by gradient flow under $F_{24}$, only partial-\textbf{kaq}s. Although there was one definitive full \textbf{kaq} local minima for a related $F_{26}$ functional based on real time evolution (recall $F_{24}$ is related to imaginary time evolution). Unfortunately, we judged $F_{26}$ as computationally too space intensive: expanding to 6th order in $c$, with the source term present leads to an explosion of diagrams. So we do not have data on whether this previously discovered \textbf{kaq} minima would evolve to an $\mfsu(8)$-\textbf{skaq} minima as $J$ is turned on. But our third finding is not merely non-detection of a $\mfsu(8)$ full \textbf{skaq} minima, we find pronounced resistance to the final tensor factoring. We assess this as follows: Given a proposed factoring $\mbbC^8= \mbbC^2 \otimes \mbbC^2 \otimes \mbbC^2$, we can consider three bipartite entropies by taking the three factors into two groups in the three possible ways. (There are also more sophisticated tripartite entanglement entropies \cite{schneeloch2020quantifying} but we do not consider them here). It is elementary that if two of the three vanish so does the third. Studies of 100,000 random vectors in $\mbbC^8$ show us that the first two entropies are nearly independent variables, but when we find a partial \textbf{skaq} decomposition, i.e., low entropy, w.r.t. $8=2\times4$, both for the most entangled $H_i$ and for $\ket{\psi}$, we find ``entropy conservation''; low entropy for the $2\times4$ factoring is generally associated with higher entropy for the opposite $4\times2$ factoring. This is illustrated in our plots for the full decomposition \cref{fig:su8g2} compared to \cref{fig:su8g} for the partial decomposition. What should we make of this? In our other experiments, entropy has no trouble simply disappearing. 

We show later a $F_{24}$ local minimum in which two of the principal axes are $2\times 4$ partially \textbf{skaq} and with full $2\times2\times2$ tensor decomposition for $J$. This mixed structure suggests three qubits with an initial trivial state and a Hamiltonian entangling two of them. This may be the first glimmers of the interaction graph briefly contemplated in the introduction; we can interpret such data as an interaction graph with three vertices and one bond. In this case, fine-grained questions come to the fore: is there a preferred type of entanglement, or entangling interaction, which is favored? Additional data will be needed to probe the emerging interactions. See our conclusion section (\cref{sec:conclusion}) for further discussion. 

 
For our findings, we located $(k,k_J)$ parameter regimes in which gradient descent had reasonable learning rates and reliably converged to local minima starting from random initial conditions. As in prior works, the perturbation parameter $1/k$ was kept small partially in the hope that the evaluated diagrams reflected the structure of the regularized integral, and partly for numerical efficiency. We found that the new source coefficient $k_J$ should be kept in the range $0.1$ to $0.2$. Smaller values are unreliable numerically since the fourth power of $k_J$ enters the diagrammatic evaluations. 

For $\mfsu(4)$, $F_{24}(g,\ket{\psi})$ was minimized with $1/k= 0.01$ or $0.02$, and $0.1 \le k_J \le 0.2$, and we found 225 local minima. For $\mfsu(8)$, as in previous studies, $1/k$ was chosen smaller as $0.001$ and $k_J$ again chosen between $0.1$ and $0.2$, and we found 282 local minima. All local minima in these ranges went into our database. We highlighted above, and present in \cref{skaqsu4,skaqsu8} a few of the more striking individual \textbf{skaq} examples. However, the case for our three findings (above) is based primarily on the collective entropy statistics derived from these two samples compared to randomly generated Gaussian Ensembles. 

\section{Loss Functions}
\subsection{Loss function to find $(g,J)$}~\\
\indent We must perform the gradient descent on $F_{24}$ in the space of metrics of fixed volume $\det g = 1$ and sources of fixed norm $||J||=1$. For this purpose, similar to earlier works \cite{freedman2021universe,freedman2021universeII,freedman2021universeIII}, we have found the Lagrangian approach to be more numerically stable instead of direct normalization by $\det(g)$ and $||J||$, and thus consider the following loss function:
\begin{align}\label{L_24}
    \mcL_{24}(g,J,k,k_J) = r_1^{-1}F_{24}(g,J,k,k_J) + r_2(\det(g)-1)^2 + 100r_2(||J||-1)^2, 
\end{align}
where $r_1 \ge 1 , r_2 \gg 1$. Gradient descent on $\mcL_{24}$ gives local minima $(g,J)$, which we call \textit{solutions to} $F_{24}$. Our solutions $g$  have highly degenerate eigenspaces, just as the solutions to the functionals in \cite{freedman2021universe} did. We review the relevant definition:
\begin{dfn}[$\text{\cite[Definition 2.1]{freedman2021universe}}$]\label{degpatdfn}
The \textit{degeneracy pattern} $(d_1,\ldots,d_t)$ is a tuple describing the dimensions of the eigenspaces ordered by increasing eigenvalues, i.e. from \textit{easier} to \textit{harder} directions.
\end{dfn}

\begin{rmk}[Representation basis]
    In this paper, as $n$ is a power of two, we choose the \emph{Pauli word} basis to express our tensors in. A Pauli word is a tensor product of the Pauli matrices and Identity, e.g.\ $\sqrt{-1} \  Z \otimes 1 \otimes X \otimes Y \otimes 1 \otimes X$\footnote{$\sqrt{-1}$ is chosen to make the word skew-Hermitian.} is a word in $\mfsu(2^6)$. This choice of default basis is consistent with the earlier works.
\end{rmk}
\begin{rmk}[Gradient descent details]
The Adam gradient descent algorithm is used (\cite[Section 2.3]{freedman2021universe}) for $\mcL_{24}$ with default hyperparameters. We choose the method \textbf{GenPerturbId} \cite[Section 2.2]{freedman2021universe} for initializing our metric, i.e. a random metric given by a Gaussian perturbation of the identity metric, where no particular \textbf{kaq} basis is favored a priori, in contrast to diagonal initializations.
\end{rmk}
Due to numerical reasons outlined in \cite{freedman2021universe}, we will not take $g$ but its inverse as our parameter for gradient descent. This reduces the number of matrix inversions that the algorithm has to compute and propagate the gradient through, enhancing numerical stabilization.

\subsection{Loss function to find the \textbf{kaq} basis}~\\
\indent The metrics of the solutions to $F_{24}$ have to go through a \textbf{kaq}ness search. We review briefly the procedure in \cite[Section 3.2]{freedman2021universeII} defining the relevant loss function.

Let the eigenbasis of $g$ be $\{iH_1, \ldots, iH_{n^2-1}\}$ with degeneracy pattern $(d_1,\ldots,d_t)$. There are two sources for the parameters $\theta$ of our loss function. The first set comes from the conjugation of the eigenbasis in $U(n)$, which is what describes the function $\mathbf{q}$ in \cref{dfn:skaq}. Further, every degenerate eigenspace of degree $d$ has an independent change of basis by an orthogonal matrix $V \in O(d)$. Thus $\dim \theta = n^2 + \sum_{i=1}^{t} (d_i^2-d_i)/2$.

After the above two transformations, we view $H_j$ as an $n^2\times1$ vector $v_{j}$, and given $n=\prod_{i=1}^l p_i$, the entropy $s_{ij}(g,\theta)$ for $1\le i \le l, 1\le j \le n^2-1$ is computed using the Schmidt decomposition of $v_j$, giving a measure of to what extent $H_j$ can be decomposed into a tensor product of two factors of size $n/p_i,p_i$ (for more details, see \cite[Section 3.2]{freedman2021universeII}). Finally, we have the \textbf{kaq} loss function defined as
\begin{align}
    \mcL_{\textbf{kaq}}(g,\theta) = \sum_{\substack{1\le i \le l \\ 1\le j \le n^2-1}} s_{ij}(g,\theta).
\end{align}

\begin{rmk}\label{partialkaqdfn}
As mentioned before, for $n=8=p_1p_2p_3=2\cdot2\cdot2$, we will investigate partial-\textbf{kaq}ness decomposition into $\C^2 \otimes \C^4$ in addition to the full qubit decomposition. This means the loss function is any of the following three: $\sum_{1\le j \le n^2-1} s_{1j}(g,\theta)$ or $\sum_{1\le j \le n^2-1} s_{2j}(g,\theta)$ or $\sum_{1\le j \le n^2-1} s_{3j}(g,\theta)$. These three along with the one for the full \textbf{kaq} decomposition give 4 loss functions for our 282 $\mfsu(8)$ solutions.
\end{rmk}
\begin{rmk}\label{partialkaqcombined}
For our plots and the discussions around partial-\textbf{kaq}ness, when we ask whether $g$ or $J$ is partial-\textbf{kaq}, we always consider the lowest entropy given by the three gradient descent searches associated to the three possible decompositions (similarly, we do so for our random baselines).
\end{rmk}
We refer to \cite[Sec. 3.2.4]{freedman2021universeII}, and \cite[Remark 3.7]{freedman2021universeII} on the gradient descent details and the threshold for deciding \textbf{kaq}ness. In brief, we take the SGD gradient descent algorithm instead of Adam due to its numerical stability when dealing with complex parameters, and we consider $\max_{i,j} s_{ij} \sim 10^{-3}$ to be the strongest indication of \textbf{kaq}ness of the metric, meaning what we would consider as \textbf{kaq} in the ideal sense if infinite precision was possible. 
\begin{rmk}\label{rmk:tolerancemargin}
We note a small change in our algorithm compared to \cite{freedman2021universeII}, where the tolerance margin for eigenvalues that are considered the same is taken to be $0.015$ instead of $0.02$. This makes the eigenspaces less degenerate and thus the \textbf{kaq}ness search harder due to the lower number of parameters in $\theta$.
\end{rmk}
\begin{rmk}
One could add the entropy of $J$, which is measured similarly to that of the principal axes of $g$ and equals $2S(\psi)$, with a suitable scalar to $\mcL_{\textbf{kaq}}$ to reach a simultaneous \textbf{kaq} basis, but simulations showed that this change of the loss function results in worse basis for both $g$ and $J$.
\end{rmk}
\vspace{1mm} 

\subsection{Parameters and data collection}~\\
\indent We always choose $r_1=1, r_2=1000$. Similar to \cite[Tables 1-2 (\textbf{GenPerturbId})]{freedman2021universe}, for $n=4$, we choose $k=100,200$ and for $n=8$, we select $k=1000$. We recall the significance of the factor $k$ in Chern-Simons theory \cite{bar1995perturbative} and in our case, in stabilizing the numerics, where smaller $\frac{1}{k}$ values are favored \cite{freedman2021universe}. For the new parameter $k_J$, we decided to take the range of $[0.1,0.125,0.15,0.175,0.2]$. We also made a few simulations with higher values of $k_J$ which showed similar patterns, but there are not in our dataset. However, going lower than $0.01$ would have made it difficult for the $J$-terms to have an impact on the value of $F_{24}$ due to the scale of the terms solely involving $c,g$ and the fourth power of $k_J$ entering the equation. 

We obtained 225 $(g,J)$ local minima for $\mfsu(4)$ and 282 for $\mfsu(8)$, with approximately the same number of simulations for pairs $(k,k_J)$ in each case. Each $g$ was then fed into the loss function $\mcL_\textbf{kaq}$, and we obtained 225 bases for $\mfsu(4)$, and $4\times 282$ (due to the four different decompositions) for $\mfsu(8)$. Lastly, we computed the entropy of $J$ w.r.t. the basis found.

\section{\textbf{skaq}ness of $F_{24}$ solutions}
As previously mentioned, once the best basis is found for $g$, we measure the entropy of $J$ w.r.t. that basis. First, we answer the question of whether there are (m)any ideal \textbf{skaq} solutions. Then we describe how we can generate random baselines for our solutions and compare them to decide whether they exhibit random or extraordinary behaviors.
\subsection{Are there (m)any ideal \textbf{skaq} solutions?}~\\
\indent Ideally, we would like to have very low entropy that signify mathematical \textbf{skaq}ness.  The criteria for an \textbf{skaq} mentioned in our summary in \cref{sec:summary} was $S(J)$ in the lower 5 percentile w.r.t. the random baseline, and the much stronger criteria for $g$: A complete set of principle axes found all with lower entropy than any from our random baseline. We have shown some examples of this ideal \textbf{skaq}ness in \cref{skaqsu4,skaqsu8}. Perhaps collecting more solutions would have given more of such examples, however those could be argued to be expected as a result of the large number of solutions collected.  However our three findings rest, not on the few very sharp \textbf{skaq} examples, but rather on the distinctive low entropy features collected from the entire ensemble of local minima in contrast with random baselines, where no such features are present.

\begin{table}[h]
    \centering
    \begin{center}
    \begin{tabular}{ |c|c|c|c| } 
    \hline
    Degeneracy pattern & Eigenvalues & Entropy ($J$) & Entropy ($g$)  \\
    \hline
    $(1, 4, 8, 2)$ & $(0.76, 0.83, 0.938, 2.12)$ & $0.0242$ & $0.0006$ \\ 
    \hline
    $(1, 3, 1, 8, 2)$ & $(0.347, 0.407,  0.448,  0.97, 10.91)$ & $0.0177$ & 1e-5\\
    \hline
    \end{tabular}
    \end{center}
    \caption{\small{\textbf{Skaq} solutions for $\mfsu(4)$. Each row is a solution $(g,J)$ with its degeneracy pattern and eigenvalues for $g$. For example, in the first solution, the eigenvalue $0.76$ is the only nondegenerate one. Both solutions have $J$ entropy in-between the 0.5 and 1 percentile of the random baseline in \cref{fig:su4J}. Note the maximum entropy of their metric principal axes passes the numerical test for \textbf{kaq}ness, meaning we would consider them to be mathematically \textbf{kaq} if infinite precision was available.}}
    \label{skaqsu4}
\end{table}
\begin{table}[h]
    \centering
    \begin{center}
    \begin{tabular}{ |p{4cm}|p{6cm}|p{2cm}|p{2cm}| } 
    \hline
    Degeneracy pattern & Eigenvalues & Entropy ($J$) & Entropy ($g$)  \\
    \hline
    (1, 16, 1, 6, 2, 16, 16, 1, 2, 2) & (0.542, 0.552, 0.6029, 0.742,
    0.775, 0.859, 1.268, 3.411, 5.687, 29.967) & $0.267$ & $0.0013$ \\ 
    \hline
    Same as above & Same as above & $0.418$ & $0.0009$\\
    \hline
    \end{tabular}
    \end{center}
    \caption{\small{Partial-\textbf{skaq} solutions for $\mfsu(8)$. The degeneracy pattern found above is quite dominant (238 out of 282) among our $\mfsu(8)$ solutions,  even though they all refer to different metrics. While the metrics are ideally \textbf{kaq}, their $J$ entropy is in the 1 and 5 percentile in \cref{fig:su8J}.}}
    \label{skaqsu8}
\end{table}

\subsection{Random baselines and findings}\label{sec:genrandombaselines}~\\
\indent To have an apple-to-apple comparison, we need to incorporate the freedoms we have in finding our \textbf{kaq} basis for $g$ into its random baseline. In other words, simply generating random $g$s and asking them to be \textbf{kaq} with respect to the standard basis would not be fair, because a random $g$ has little spectral degeneracy and therefore fewer search parameters for low entropy eigenvectors. So to construct the random baseline we input our observed degeneracy patterns, which increases the number of parameters at $\mcL_\textbf{kaq}$'s disposal to find a \textbf{kaq} basis. 

Therefore, given $M$ many solutions (recall $M=225$ or 282), we generate random metrics $g$ with the same distribution of degeneracy patterns. To do so, we simply pick a diagonal matrix $D$ with $\det D = 1$ and random eigenvalues having the desired degeneracy pattern, and conjugate it by a random orthogonal matrix of the same size which is itself generated using a random skew-symmetric matrix which entries are sampled from a Gaussian distribution with mean zero and variance 1/10. For fairness, we have chosen this small variance since our own solutions have roughly this variance and generating orthogonal matrices further from identity is prone to produce worse \textbf{kaq} results. Then we run the \textbf{kaq} gradient descent to find the best basis for each random $g$. Finally, we plot the maximum entropy among the principal axes.

\begin{figure}[h]
    \centering
    \includegraphics[scale=0.6]{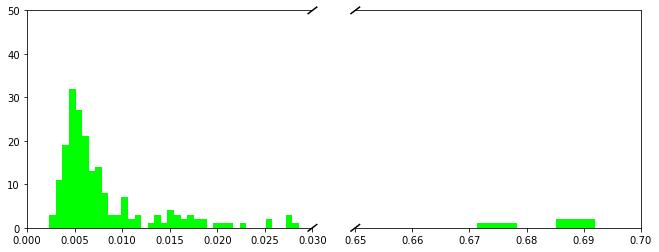} \\
    \includegraphics[scale= 0.6]{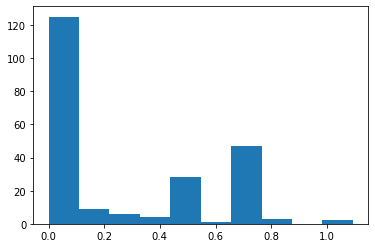}
    \includegraphics[scale= 0.6]{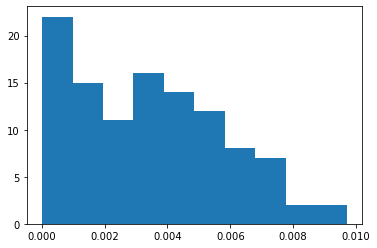}
    \caption{\small{\textbf{Top}: Histogram plot generated by the maximum entropy of the principal axes of 225 random $g$s. There are a few outliers in the data, hence why we have used an interrupted $x$-axis. Notice how the majority (197 out of the 225) of the entries are present in the left subplot. \textbf{Bottom left}: We plot the maximum entropy of the principal axes of $g$ in the 225 solutions for $\mfsu(4)$. Similar to the subplot on top left, we see a stronger presence on the left side, with the leftmost bin containing $>120$ of the solutions. \textbf{Bottom right}: We zoom in on that leftmost bin, more precisely all entries below $0.01$, and observe a very different pattern from the top left subplot: There is a much stronger concentration of low entropies with 37 entries out of the total 225 below 2e-3. There is no entry below that value in the random baseline.}}
    \label{fig:su4g}
\end{figure}
\begin{figure}[h]
    \centering
    \includegraphics[scale = .6]{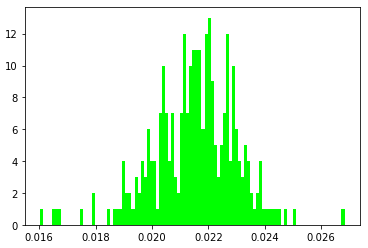}\\
    \includegraphics[scale= .6]{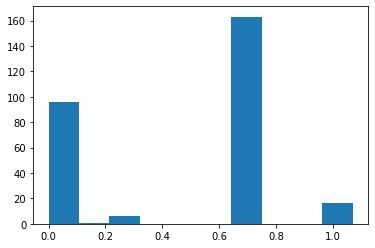}
    \includegraphics[scale= .6]{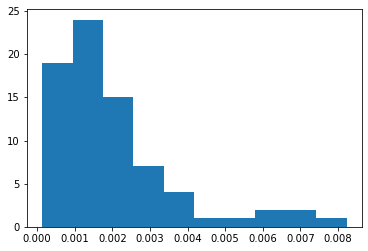}
    \caption{\small{\textbf{Top}: Histogram plot with 100 bins generated by the maximum entropy of the principal axes of 282 random $g$s with respect to the $\mbbC^2 \otimes\mbbC^4$ decomposition in $\mfsu(8)$. \textbf{Bottom left}: Histogram plot with 10 bins of the maximum entropy of principal axes of $g$ in the 282 solutions of $\mfsu(8)$ w.r.t. the same decomposition. We note the strong tick on the left, but also the one on the right which are entries $\sim$ 0.69. It should be noted that the \textbf{kaq} search in those cases has settled on a basis that has lowered the sum of all entropies, and has managed to make the majority of principal axes disentangled but has left a few of the principal axes with high entropy. We hypothesize that the reason why some of the bins are empty is due to some common structure that the local minima of $F_{24}$ share, so that even though they are different local minima, they often land on bases that provide, to some approximation, the same maximum entropy among the principal axes. \textbf{Bottom right}: We zoom in on the leftmost bin of the previous plot with entries below $0.012$, and note not only a distribution skewed to the left, but also the fact that all $76$ entries in this histogram are not captured in the random baseline on top.}}
    \label{fig:su8g}
\end{figure}
\begin{figure}[h]
    \centering
    \includegraphics[scale = 0.6]{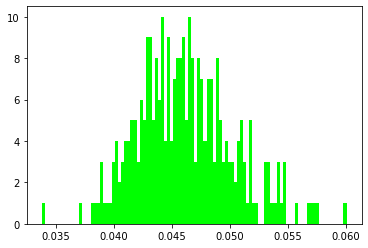}
    \includegraphics[scale= 0.6]{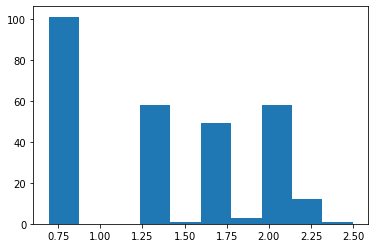}
    \caption{\small{\textbf{Left}: Histogram plot with 100 bins generated by the maximum entropy of the principal axes of 282 random $g$s with respect to the $\mbbC^2 \otimes \mbbC^2 \otimes\mbbC^2$ decomposition in $\mfsu(8)$. Notice how the scaling on the $x$-axis is about twice the partial decomposition plot in \cref{fig:su8g}, which is the result of adding the entropy of two partial decompositions $2\times 4$ and $4\times 2$. \textbf{Right}: 10 bins histogram plot of the maximum entropy of principal axes of $g$ in the 282 solutions of $\mfsu(8)$ w.r.t. the same decomposition. Notice that even the scale on the $x$-axis do not match that of the random baseline. All our data is located to the extreme right of the random baseline and one may check that the explanation in \cref{fig:su8g} for the similar situation does not apply, namely the \textbf{kaq} search has not even managed to truly lower the sum of all entropies and there are quite many of the principal axes with high entropies. As such, we can conclude a strong resistance to full decomposition.}}
    \label{fig:su8g2}
\end{figure}

We obtain one histogram plot for $\mfsu(4)$ (\cref{fig:su4g}) and four for $\mfsu(8)$, and note that three of these (for partial-\textbf{kaq}) are similar but have to be considered together (\cref{partialkaqcombined}), hence why we have plotted two kinds of histograms for $\mfsu(8)$ (\cref{fig:su8g,fig:su8g2}). All random baselines (colored green) are compared within the same figure to their counterpart (colored blue) among our metric solutions to $F_{24}$. 

\begin{figure}[h]
    \centering
    \includegraphics[width=.48\textwidth]{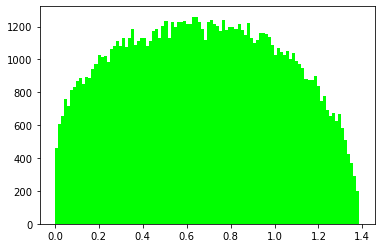}
    \includegraphics[width=.48\textwidth]{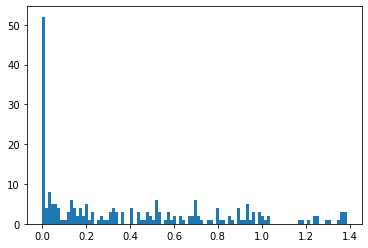}
    \caption{\small{\textbf{Left}: Histogram plot on 100,000 random samples for the entropy of $J$ in $\mfsu(4)$. The number of bins used is 100. We note the range of $[1.75\text{e-}5,1.39]$. The $0.1,0.5,1,5,10,25,50$ percentiles are  $0.0036, 0.0148, 0.0262, 0.0995, 0.1779, 0.3725, 0.6645$, respectively. The mean and variance are $0.668,0.1279$. \textbf{Right}: Histogram plot with 100 bins on the entropy of $J$ in the 252 solutions for $\mfsu(4)$. We observe a strong signal of low entropy compared to the random baseline where 56 and 79 entries are below the 1 and 5 percentile, respectively.}}
    \label{fig:su4J}
\end{figure}

\begin{figure}[h]
    \centering
    \includegraphics[width=.48\textwidth]{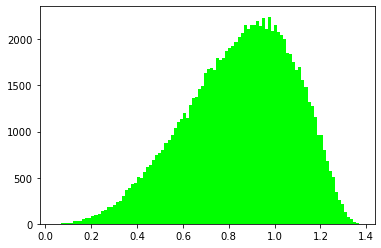}
    \includegraphics[width=.48\textwidth]{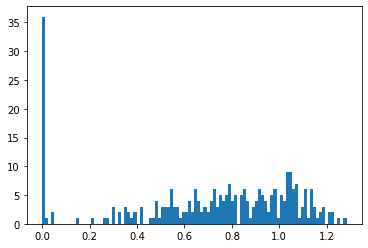}
    \caption{\small{\textbf{Left}: Histogram plot on 100,000 random samples for the entropy of $J$ with respect to the $\mbbC^2 \otimes \mbbC^4$ decomposition in $\mfsu(8)$. The number of bins used is 100. We note the range of $[0.0417, 1.3693]$. The $0.1,0.5,1,5,10,25,50$ percentiles are $0.1527, 0.2312, 0.277, 0.4303, 0.5247, 0.6919, 0.8715$, respectively. The mean and variance are $0.8488, 0.0538$. \textbf{Right}: This plots the entropy of $J$ in 100 bins with respect to the $\mbbC^2 \otimes \mbbC^4$ decomposition for the 282 solutions found for $\mfsu(8)$. We note the extreme unlikeliness of the strong signal on the leftmost bin when compared to the random plot: All entries below 0.2 are in the 0.5 percentile of the random baseline, and there are 39 such entries.}}
    \label{fig:su8J}
\end{figure}

\begin{figure}[h]
    \centering
    \includegraphics[width=.48\textwidth]{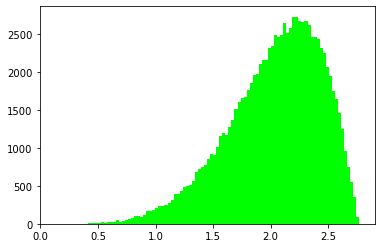}
    \includegraphics[width=.48\textwidth]{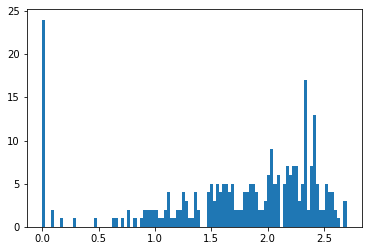}
    \caption{\small{\textbf{Left}: Histogram plot on 100,000 random samples for the entropy of $J$ with respect to the $\mbbC^2 \otimes \mbbC^2 \otimes\mbbC^2$ decomposition in $\mfsu(8)$. The number of bins used is 100. We note the range of $[0.1255, 2.7689]$. The $0.1,0.5,1,5,10,25,50$ percentiles are  $0.6074, 0.8222, 0.9365, 1.2822, 1.4742, 1.7841, 2.0923$, respectively. The mean and variance are $2.0356, 0.1618$. \textbf{Right}: Similar to the partial decomposition, and in contrast with \cref{fig:su8g2}, we observe an extremely unlikely event for the entropy of $J$ w.r.t. the $\mbbC^2 \otimes \mbbC^2 \otimes\mbbC^2$ decomposition, where there are 29 entries out of 282 below the 0.1 percentile of the random baseline.}}
    \label{fig:su8J2}
\end{figure}

The random baseline for $J$ is easier to make than that of $g$, since there is no degeneracy pattern or gradient descent to be taken into account. Therefore, we can simply generate random vectors $J$ on $S^7$ ($S^{15}$) in $\mbbC^4$ ($\mbbC^8$), and see if their corresponding $\bfJ$ in $\mfsu(4)$ ($\mfsu(8)$)  is a \textbf{kaq} operator in any fixed basis like our default Pauli word basis. Similar to the previous case, we obtain one plot for $\mfsu(4)$ (\cref{fig:su4J}) and two for $\mfsu(8)$ (\cref{fig:su8J,fig:su8J2}). Furthermore, one can generate many more samples than we did for the metric; we generate 100,000 for each random baseline. Again, the random baselines and their counterpart among our $J$ solutions to $F_{24}$ are compared in the same figures.

We have discussed in the caption of each figure the significance of our findings and we refer to our summary of results (\cref{sec:summary}) for the outline.

\begin{figure}
    \centering
    \includegraphics[width=.96\textwidth]{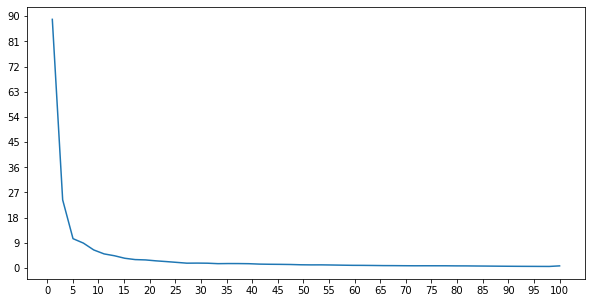}
    \caption{\small{\textbf{Skaq} percentiles comparison to random baseline for $\mfsu(4)$. $x$-axis denotes the percentile and $y$-axis the ratio of the percentage of solutions (out of 225) in that percentile over $(q\%)^2$. We note that in low percentile $1-5$ territory, our data exhibits a strong \textbf{skaq} signal by having $\sim90-9$ times relatively more entries than the random baseline.}}
    \label{fig:su4skaqpercentiles}
\end{figure}

\begin{figure}[h]
    \centering
    \includegraphics[width=.96\textwidth]{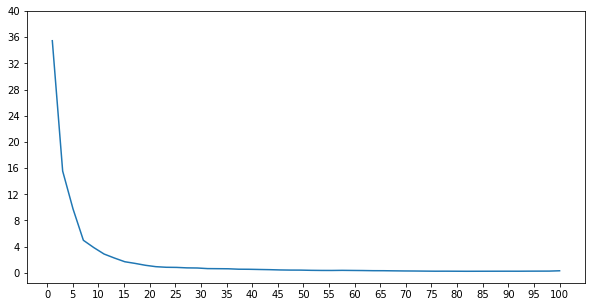}
    \caption{\small{\textbf{Skaq} percentiles comparison to random baseline for the $\mbbC^2 \otimes\mbbC^4$ decomposition for $\mfsu(8)$. $x$-axis denotes the percentile q and $y$-axis the ratio of the percentage of solutions (out of 282) in that percentile over $(q\%)^2$. Again, we see that in low percentile $1-5$ territory, our data exhibits a strong \textbf{skaq} signal by having $\sim35-10$ times relatively more entries than the random baseline.}}
    \label{fig:su8skaqpercentiles}
\end{figure}

The \cref{fig:su4skaqpercentiles,fig:su8skaqpercentiles} bring our results on $g$ and $J$ together to discuss \textbf{skaq}. To make a joint comparison to the random baselines, we will need to take a percentile $q$ and compute the corresponding thresholds determined by the random baselines of $g,J$, calling them $p_g, p_J$. Then we compute the percentage of our $(g,J)$ solutions that had their respective entropy smaller than $p_g$ and $p_J$. Finally, we divide that amount by $(q\%)^2$. The larger this amount is from 1, the more unlikely it is that our data is random. We see a strong signal in the low $1-5$ percentile territory which is the region most important to us.

\FloatBarrier

\section{Conclusions, mysteries and outlook}\label{sec:conclusion}
Earlier work \cite{freedman2021universe,freedman2021universeII,freedman2021universeIII} studied symmetry breaking in the space of metrics, with the metrics themselves being placed on the space of Hamiltonians. The earlier work found that the spontaneously-determined metrics naturally give rise to qubit (or qunit) structures  (called \textbf{kaq}). The qubit structure on the metrics led to qubit structures in the Hamiltonian of the universe:  in the low temperature limit, the  Hamiltonian is given by the metric's shortest principal axis. In this paper, we have sharpened the question to study spontaneous symmetry breaking to pairs of \{metric, initial quantum state\}. If such pairs simultaneously know about the same qubit structure they are called \textbf{skaq}. Broadly, we find that the universe still likes qubits, with two provisos. First, the initial state is generally less tightly aligned with the optimal qubit structure than the metric, but still unmistakably aligned. Second, in our study of $\mfsu(8)$, that is on $\C^8$, we found a bug, which we can also see as a feature. The ``bug'' was the initially disappointing result that we see $\C^8$ decompose as $\C^2 \otimes \C^4$, $8=2\times 4$; but we do not see $\C^8$ decompose all the way into $\C^2 \otimes \C^2 \otimes \C^2$, $8=2\times 2\times 2$, in the \textbf{skaq} context. In fact we see a rather extreme resistance to this final factoring evidenced by the tallest bar in the lower left plot of \cref{fig:su8g} and the right plot of \cref{fig:su8g2}. Both bars represent a profound resistance to a final factorization of the metric, evidenced by the most resistant principal value.  Much of them (all in that bar in \cref{fig:su8g}, and 83 out of 101 in that bar in \cref{fig:su8g2}) concentrate at a 0.005 distance around a specific ``mystery'' value, 0.69. We suspect there is some highly entangling dynamic, perhaps a well-known form of interaction, producing it (a future investigation?). Looking at our data on the tensor decomposition of eigenvectors of $g$ into operators acting on $\C^2\otimes\C^4$, we can exclude the case of Heisenberg interaction for the matrix factors acting on $\C^4$, and see many of them that have a spectrum with degeneracy patterns such as $(2,1,1)$ or $(2,2)$, and eigenvalues very close to $\frac{\pm1}{\sqrt{2}}$ and $0$, when the operators are normalized. For the latter type of eigenvectors, we showcase two of them in \cref{eq:eigenvecs_entangling} with their respective 2-dimensional subspace in $\C^4$ (that is orthogonal to their kernel) generated by the column vectors. We note that the left one in \cref{eq:eigenvecs_entangling} is from the cheapest eigen-direction and the right one from one of the most expensive ones, both written in a basis in which the $4$-d tensor factor of $J$ factors into $\C^2\otimes \C^2$. These $2$-d subspaces are where the entanglement happens. These entangling interactions are reproduced in the hope that some reader may recognize them as something familiar from condensed matter or quantum information; the authors have no insight into these matrices.

\begin{align}\label{eq:eigenvecs_entangling}
  \begin{pmatrix}
 -0.2137-0.5499i &  0.0075+0.2251i \\
-0.4282+0.1446i &  0.3625-0.2245i \\
0.6284+0.0000i & 0.1901-0.0489i \\
0.2268-0.0363i &  0.8538+0.0000i
\end{pmatrix} , \ 
\begin{pmatrix}
-0.2164-0.5494i & 0.0122+0.2234i \\
-0.4260+0.1484i &  0.3609-0.2286i \\ 
0.6289+0.0000i  & 0.1901-0.0445i \\  
0.2266-0.0302i  & 0.8540+0.0000i 
\end{pmatrix}
\end{align}

What we know for sure is that once 8 splits as $2\times4$ there is more than random resistance, on the part of the metric to further decompose that 4 into $2\times2$. But now comparing those two bars in \cref{fig:su8g,fig:su8g2} with \cref{fig:su8J2}, the ``bug'' starts to look like a feature. Figure~\ref{fig:su8J2}, which discuss only $J$ ($J$ encodes the initial vector),  shows in the data that it quite often decomposes as a triple tensor product (evidenced by the entropy spike at zero). So what we may be looking at, is an initially disentangled state in three qubits, and an initial Hamiltonian which is the sum of two terms: 1-body on qubit 1, and some entangling interaction between qubits 2 and 3. Now this is starting to look like a ``feature''. It is the faintest glimmer of Greg Moore's suggestion (see \cref{ftnote:gregmoore}) that perhaps space might emerge from our toy model as a lattice of qubits with bonds representing strong pairwise interactions. Greg was hoping that a good chunk of the Leech lattice might arise in a large version of our toy model. More modestly, what we (may) have before us is three vertices and one bond joining two of them. We find this intriguing result heightens our impatience for a $10^{11}$ logical-qubit quantum computer, and also our interest in complementing our (classical) numerical study with some analytic one that can give insight into higher dimensions without relying on brute force.

As a final thought, we have explored spontaneous symmetry breaking at the level of operators (technically metrics whose principal axes are operators) but could there be a way to transcribe this discussion of operator symmetry breaking into the more familiar context of spontaneous symmetry breaking of states? After all, an experiment in an atomic physics lab that creates an effective Hamiltonian for some system does not do so by changing the Standard Model of particle physics; instead, working always within the context of the base layer of laws of physics, it changes the effective Hamiltonian by changing the \textit{state} of the lab-equipment to which the system is coupled. Can we analyse  spontaneous symmetry breaking in the space of Hamiltonians, or the space of states and Hamiltonians, in the context of regular spontaneous symmetry breaking in the space of states in this fictitious extended Hilbert space? Indeed, is there some way to realize this extended Hilbert space in a lab, giving an experimental probe of systems far larger than the smallest interacting universe?

\section*{Acknowledgments}
 The last author would like to thank the Aspen Center for Physics for their hospitality and research environment. The experiments were conducted using Microsoft computational resources.

\bibliographystyle{apa}
\bibliography{main}

\end{document}